\documentclass[10pt,twocolumn]{article} 
\usepackage{simpleConference}
\usepackage{times,natbib}
\usepackage{graphicx}
\usepackage{amssymb}
\usepackage{url,hyperref}
\usepackage{cite}
\usepackage{amsmath}
\usepackage{tensor}
\usepackage{booktabs}
\DeclareMathOperator{\Tr}{Tr}

\begin{document}

\title{Smoothing and Interpolating Noisy GPS Data with Smoothing Splines}
\author{Jeffrey J. Early, Northwest Research Associates, USA \\ Adam M. Sykulski, Lancaster University, UK}
\maketitle
\thispagestyle{empty}

\abstract{A comprehensive methodology is provided for smoothing noisy, irregularly sampled data with non-Gaussian noise using smoothing splines. We demonstrate how the spline order and tension parameter can be chosen \emph{a priori} from physical reasoning. We also show how to allow for non-Gaussian noise and outliers which are typical in GPS signals. We demonstrate the effectiveness of our methods on GPS trajectory data obtained from oceanographic floating instruments known as drifters.}
\\ \\
\noindent This work has not yet been peer-reviewed and is provided by the contributing author(s) as a means to ensure timely dissemination of scholarly and technical work on a noncommercial basis. Copyright and all rights therein are maintained by the author(s) or by other copyright owners. It is understood that all persons copying this information will adhere to the terms and constraints invoked by each author's copyright. This work may not be reposted without explicit permission of the copyright owner.


%
\section{Introduction}
%
In the summer of 2011 an array of floating ocean surface buoys (drifters) were deployed in the Sargasso Sea to assess the lateral diffusivity of oceanic processes \citep{shcherbina2015-bams}. Each drifter was equipped with a global positioning system (GPS) receiver recording locations every 30 minutes.
Addressing the primary goal of understanding the physical processes controlling lateral diffusivity requires significant processing of the drifter positions, including removing the mean flow across all drifters, accounting for the large scale strain field, and analyzing the residual spectra for hints of a dynamical process. However, it quickly became clear that the GPS position data, which can have accurracies as low as a few meters \citep{faa2016-report}, was contaminated by outliers with position jumps of hundreds of meters or more. Prior to analysis, the position data requires removing the outliers, and interpolating gaps to keep the position data synchronized in time across the drifter array.

The basic problem is ubiquitous: observations from GPS receivers return observed positions $x_i$ at times $t_i$ that differ from the true positions $x_{\textrm{true}}(t_i)$ by some noise $\epsilon_i \equiv x_i - x_{\textrm{true}}(t_i)$ with variance $\sigma^2$. The primary goal of \emph{smoothing} is to find the true position $x_{\textrm{true}}(t_i)$ not contaminated by the noise, while the primary goal of \emph{interpolating} is to find the true position $x_{\textrm{true}}(t)$ between observation times.

The approach taken here is to use smoothing splines. Our model for the `true' path $x(t)$ is specified using interpolating b-splines $X^K(t)$ such that
\begin{equation}
\label{b-spline-model-intro}
    x(t) = \sum_{i=1}^N \xi_i X^K_i(t),
\end{equation}
where $K$ is the order (degree $S=K-1$) of the spline. For $N$ observations we construct $N$ b-splines such that $x(t_i)=x_i$ for appropriately chosen coefficients $\xi_i$. To smooth the data we choose new coefficients $\bar{\xi_i}$ that minimize the penalty function
\begin{equation}
\label{smoothing-spline}
\phi =  \frac{1}{N}\sum_{i=1}^{N} \left( \frac{x_i - x(t_i)}{\sigma} \right) ^2 + \frac{\lambda_T}{t_N-t_1} \int_{t_1}^{t_N} \left(\frac{d^T x}{dt^T}\right)^2 \, dt,
\end{equation}
for some tension parameter $\lambda_T \geq 0$. If $\lambda_T = 0$ then $\phi=0$ and $\xi_i=\bar{\xi_i}$ because $x(t_i)=x_i$, but if $\lambda_T \rightarrow \infty$ then this forces the $x(t)$ to a $T$-th order polynomial (e.g., when $T=2$, the model is forced to be a straight line because it has no second derivative). The resulting path $x(t)$ is known as a smoothing spline and was first introduced in modern form by \citet{reinsch1967-nm}, but according to \citet{deboor1978-book} the idea dates back to \citet{whittaker1923-pems}. Once $S$ and $T$ are chosen, the smoothing spline has one free parameter ($\lambda_T$) and its optimal value can be found by minimizing the expected mean square error when the true value of $\sigma$ is known \citep{craven1979-nm}.

As a practical matter there are three issues that must be addressed before smoothing splines are applied to GPS data:
\begin{enumerate}
    \item how do we choose $S$ and $T$---and how do these choices affect the recovered power spectrum?
    \item how do we modify the spline fit to accommodate the non-Gaussian errors of GPS receivers?
    \item how do we identify and remove outliers?
\end{enumerate}
To address these issues, but also serve as a practical guide to other practitioners, we start by reviewing B-splines in section \ref{sec:interpolation} and introduce the canonical interpolating spline that is used as the underlying model for path $x(t)$ in \eqref{b-spline-model-intro}. We also demonstrate the effect that choosing $S$ has on the high-frequency slope of the power spectrum of the interpolated fit.

Section \ref{sec:smoothing_spline} takes a broad look at smoothing splines and the assumptions they make on the underlying process. Many of the ideas presented in this section are known to the statistics community, so here we present these ideas from a more physical perspective. 
We show that the penalty function in \eqref{smoothing-spline} can be formulated as a maximum likelihood problem and applying tension is equivalent to assuming a Gaussian distribution on the tensioned derivative of the underlying process.

Section \ref{sec:spline_order_tension_order_spectrum} uses ensembles from synthetic data designed to mimic the oceanographic data in order to test a number of choices that have to be made. We first establish that setting $T=S$ is a reasonable choice. We then show that the tension parameter can be chosen \emph{a priori} (without optimization of the mean square error) when the {\em effective sample size} (which we define later) can be estimated from the data. This estimate for the effective sample size can then be used to reduce the coefficients, $\xi^i$, in the spline fit without increasing mean square error. Finally, we show how the effective sample size of the fit establishes the highest resolved frequency.

The second half of the manuscript addresses issues specific to GPS positions errors. In section \ref{sec:bivariate} we discuss the assumptions of stationarity and isotropy required for bivariate smoothing splines. In section \ref{sec:drifter_data_set} we show that the GPS errors are not Gaussian distributed, but $t$-distributed, and we show how to modify the technique for a $t$-distribution. Finally, section \ref{sec:outliers} addresses how to modify the expected mean square error minimizer to make smoothing splines robust to outliers.

One of the major outcomes of this work is the implementation of Matlab classes for generating b-splines, interpolating splines, smoothing splines as well as a class specific to smoothing GPS data\footnote{https://github.com/JeffreyEarly/GLNumericalModelingKit}. These classes are highlighted throughout the manuscript in their relevant sections.

%
\section{Interpolating Spline}
\label{sec:interpolation}
%

\begin{figure*}[h]
  \centerline{\includegraphics[width=39pc,angle=0]{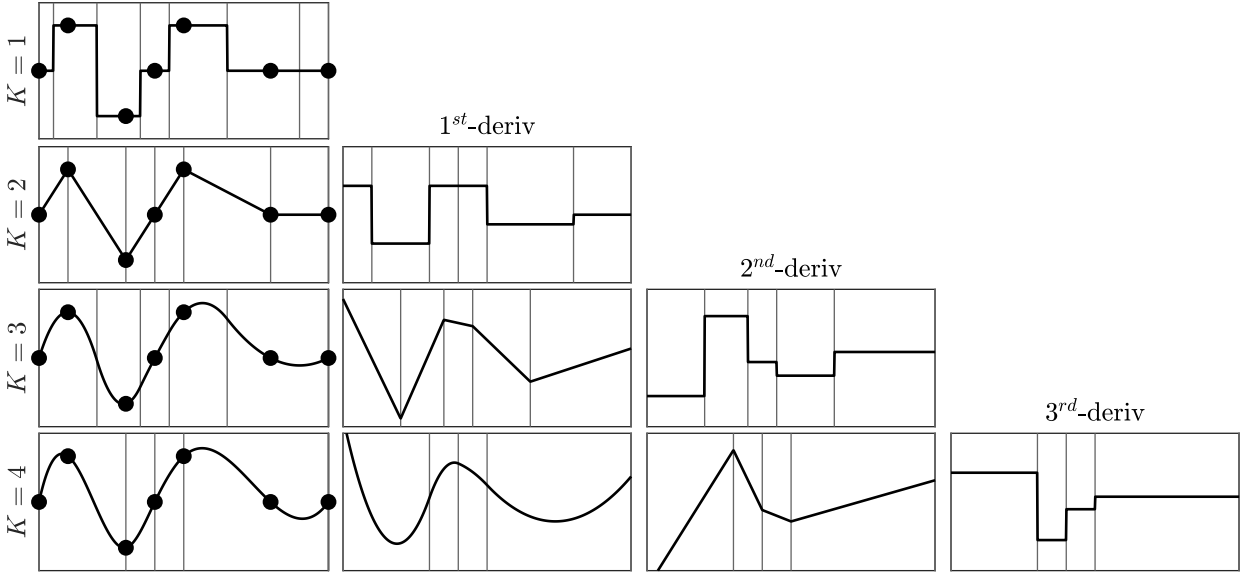}}
  \caption{An example of interpolating between 7 data points. The data points are shown as circles, and the interpolated function is shown as solid black lines. We show four different orders of interpolation $K=1..4$ (rows) and their nonzero derivatives (columns). The thin vertical grey lines are the knot points.}
  \label{interpolation}
\end{figure*}

Assume that we are given $N$ observations of a particle position $(t_i,x_i)$ with no errors. The simplest possible form of interpolation would be a nearest neighbor method that assigns the position of the particle to the nearest observations in time. The resulting interpolated function $x(t)$ is a polynomial of \emph{order} $K=1$ (piecewise constant), shown in the top row of Fig. \ref{interpolation}. The next level of sophistication is to assume a constant velocity between any two observations and use that to interpolate positions between observations, second row of Fig. \ref{interpolation}. This also means that we now have a piecewise constant function $\frac{dx}{dt}$ that represents the velocity of the particle, shown in the second row, second column of Fig.  \ref{interpolation}. This is a polynomial function of order $K=2$.

It is slightly less obvious how to proceed to a polynomial of order $K=3$. With $N$ data points we can construct a piecewise constant acceleration (the second derivative) using the $N-2$ independent accelerations computed from finite differencing, but where to place $\emph{knot points}$ that define the boundaries of the regions and how to maintain continuity is slightly less clear. The approach taken here is to use B-splines.

\subsection{B-Splines}

A B-spline (or basis spline) of \emph{order} $K$ (\emph{degree} $S=K-1$) is a piecewise polynomial that maintains nonzero continuity across $S$ knot points. The knot points are a nondecreasing collection of points in time that we will denote with $\tau_i$. The basic theory is well documented in \citet{deboor1978-book}, but here we will present a reduced version specifically tailored to our needs.

\begin{figure}
  \centerline{\includegraphics[width=19pc,angle=0]{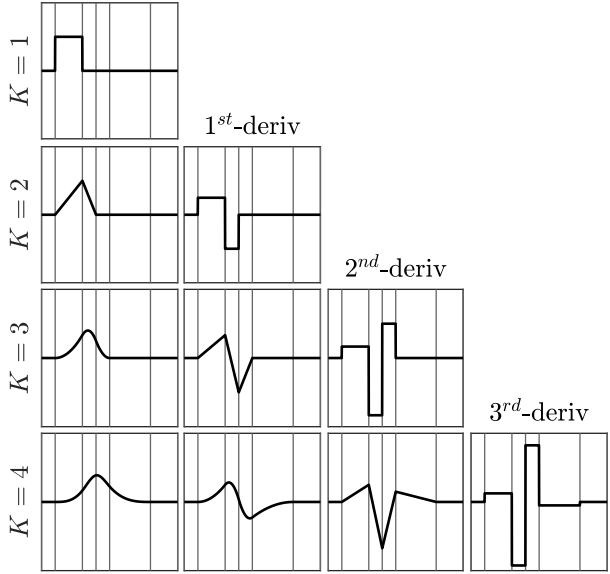}}
  \caption{B-splines and derivatives (columns) for orders $K=1..4$ (rows).}
  \label{bsplines}
\end{figure}

The $m$-th B-spline of order $K=1$ is defined as
\begin{equation}
X^1_m(t) \equiv \begin{cases}
1      & \text{if $ \tau_m \leq t < \tau_{m+1}$}, \\
0     & \text{otherwise}.
\end{cases}
\end{equation}
This is the rectangle function as shown in the first row, first column of Fig. \ref{bsplines}. If we are given $P$ knot points, then we can construct $P-1$ B-splines of order $K=1$, although notice that if a knot point is repeated this will result in a spline that is zero everywhere. To represent an interpolating function $x(t)$ for the $N$ observations of a particle position $(t_i,x_i)$ we define $N+1$ knot points as
\begin{equation}
\tau_m = \begin{cases}
t_1      & \text{$m=1$}, \\
t_{m-1} + \frac{t_m-t_{m-1}}{2}	  & \text{$1<m \leq N$},\\
t_N     & \text{$m>N$}.
\end{cases}
\end{equation}
This will create $N$ independent basis functions that provide support for the region $t_1 \leq t \leq t_N$ (provided the last spline is defined to include the last knot point). The interpolating function $x(t)$ is defined as $x(t) \equiv  X^1_m(t) \xi^m$ where the coefficients $\xi^m$ are found by solving $X^1_m(t^i) \xi^m = x^i$. The result of this process is shown in Fig. \ref{interpolation} for 7 irregularly spaced data points.

All higher order B-splines are defined by recursion,
\begin{equation}
X^K_m(t) \equiv \frac{t - t_m}{t_{m+K-1} - t_m} X^{K-1}_m(t) + \frac{t_{m+K}-t}{t_{m+K} - t_{m+1}} X^{K-1}_{m+1}(t).
\end{equation}
This recursion formula takes two neighboring lower order splines and ramps the left one up over its nonzero domain and ramps the right one down over its nonzero domain. The result of this process is to create splines that span across one additional knot point at each order, and maintain continuity across one more derivative. Examples are shown in Fig. \ref{bsplines}.

Any knot points that are repeated $T$ times will result in a total of $T-1$ splines of order one that are everywhere zero. This has the effect of introducing discontinuities in the derivatives for higher order splines. For our purposes, we will only use this feature to prevent higher order splines from crossing the boundaries. For $K=2$ order splines we will use $N+2$ knot points at locations
\begin{equation}
\tau_m = \begin{cases}
t_1      	& \text{$m \leq 2$}, \\
t_{m-1}	& \text{$2 < m \leq N$},\\
t_N 		& \text{$m > N$}.
\end{cases}
\end{equation}
This creates a knot point at every observation point, but repeats the first and last knot point. This has the effect of terminating the first and last spline at the boundary and creating $N$ second order B-splines, $X^2_m(t)$. Once again the interpolating function $x(t)$ is defined as $x(t) \equiv  X^2_m(t) \xi^m$ where the coefficients $\xi^m$ are found by solving $X^2_m(t^i) \xi^m = x^i$. The second row of Fig. \ref{interpolation} shows an example.

This process can be continued to higher and higher order B-splines. For splines that are of \emph{even} order, we create $N+K$ knots points with
\begin{equation}
\tau_m^{\text{$K$-even}} = \begin{cases}
t_1      	& \text{$m \leq K$}, \\
t_{m-K/2}	& \text{$K < m \leq N$},\\
t_N 		& \text{$m > N$},
\end{cases}
\label{even-knots}
\end{equation}
and for splines that are \emph{odd} order, we create $N+K$ knot points with
\begin{equation}
\tau_m^{\text{$K$-odd}} = \begin{cases}
t_1      	& \text{$m \leq K$}, \\
t_{m-\frac{K+1}{2}} + \frac{t_{m+1-\frac{K+1}{2}}-t_{m-\frac{K+1}{2}}}{2}	& \text{$K < m \leq N$},\\
t_N 		& \text{$m > N$}.
\end{cases}
\label{odd-knots}
\end{equation}
The knot points are chosen specifically to create $N$ splines for the $N$ data points such that the interpolated function $x(t)$ crosses all $N$ observations $(t_i,x_i)$. The path $x(t)$ is the \emph{canonical interpolating spline of order K}. Examples are shown in Fig. \ref{interpolation}.

The knot placements in \eqref{even-knots} and \eqref{odd-knots} are equivalent to the \textit{not-a-knot} boundary conditions described in \citet{deboor1978-book} and used in the cubic spline implementation in \texttt{Matlab}. In the usual formulation of the not-a-knot boundary condition, the knot positions do not change as a function of spline order, and therefore additional constraints have to be added at each order---especially the requirement that the highest derivative maintain continuity near the boundaries. In the formulation here, these constraints are implicit in  \eqref{even-knots} and \eqref{odd-knots}.


\subsection{Numerical implementation}

The root class in our suite of Matlab classes is the \texttt{BSpline} class, which evaluates a complete B-spline basis set given a set of knot points. This class was used to generate Fig. \ref{bsplines}.

The interpolating spline used to generate Fig. \ref{interpolation} is implemented in the \texttt{InterpolatingSpline} class---a sublcass of \texttt{BSpline}. This class generates interpolating splines of arbitrary order given a set of data points $(t_i, x_i)$, thus generalizing the cubic spline command that is built in to Matlab.

\subsection{Synthetic Data}
\label{sec:synthetic_data}

Throughout this manuscript we generate synthetic data for both the signal and the noise. The velocity of the signal is generated from a Gaussian process known as the Mat\'ern \citep{lilly2017-npg}. The spectrum of the Mat\'ern is given by
\begin{equation}
S(\omega) = \frac{A^2}{(\omega^2 + \lambda^2)^{p/2}},
\end{equation}
with $p>1$, which has finite amplitude at low frequencies and power-law fall off at high frequencies, two physically realistic properties observed, among other things, in ocean surface drifters \citep{sykulski2016-jrssc}.


For these experiments we choose values of $p=2,3,4$ so that the high frequency spectrum is proportional to $\omega^{-2}$, $\omega^{-3}$, $\omega^{-4}$. The Mat\'ern is used to generate the \emph{velocity} of the signal and integrated to get positions. Parameters are chosen such that the square root of velocity variance in each direction is $u_{\textrm{rms}}=0.20$ m/s and the damping scale $\lambda^{-1}=30$ minutes. These choices resemble the data from the drifters. Fig. \ref{varied_slope} shows an example velocity spectrum of the signal with $\omega^{-2}$.

The position data is contaminated with (white) Gaussian noise with $\sigma=10$ meters, a value chosen to resemble GPS errors. In section \ref{gps_position_errors} we consider noise generated from a $t$-distribution which more accurately reflects GPS errors.

For all of these experiments we use a range of \emph{strides}, that is, subsampled versions of the underlying process as input into the spline fits. A stride of 100 indicates that the signal is subsampled to 1 every 100 data points. This lets us evaluate the quality of fit against different sampling rates.

\subsection{Spline degree, $S$} \label{spline_degree}

\begin{figure}
  \centerline{\includegraphics[width=19pc,angle=0]{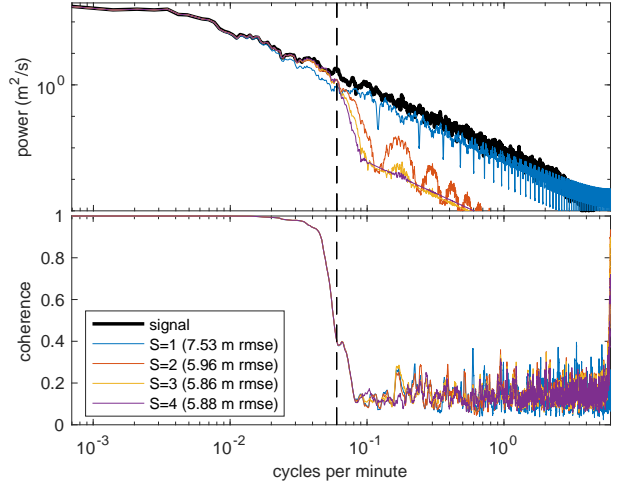}}
  \caption{The upper panel shows the velocity spectrum of the signal (black). The blue, red, and orange lines show the spectrum of the interpolating spline fit to the data with a stride of 100 for $S=1..4$, respectively. The dashed vertical line denotes the Nyquist frequency of the strided data. The bottom panel shows the coherence between the smoothed signals and the true signal.}
  \label{varied_slope}
\end{figure}

We first examine a synthetic signal \emph{uncontaminated} by noise, to examine the role of spline degree, $S$, on the interpolated fit. As noted in \citet{craven1979-nm}, the degree of the spline sets its roughness. In terms of the power spectrum, this corresponds to the high frequency slope as can be seen in Fig. \ref{varied_slope} which shows fits with $S=1..4$. Setting $S=1$ produces a high frequency fall off in the spline fit of $\omega^{-2}$. Although this would appear to be a desirable feature when fitting to a process with slope $\omega^{-2}$, the mean square error is consistently higher.

The bottom panel of Fig. \ref{varied_slope} shows the coherence between the spline fit and the true signal. There is no discernible difference in coherence between spline fits with $S=1..4$. The coherence quickly drops to near zero at the same frequency in all three cases. The implication here is that the spline fits are essentially producing noise at frequencies above the loss-of-coherence. This is why the shallower slopes (with more variance at high, incoherent frequencies) have a larger mean square error than the steeper slopes (with less variance at high, incoherent frequencies). The conclusion here is that smoother is better: it is better to use an unnecessarily high order spline to avoid adding extra noise at high frequencies.

%
\section{Smoothing Spline}
\label{sec:smoothing_spline}
%

A typical starting point for maximum likelihood is to establish the probability distribution function (PDF) of the errors, $\epsilon_i \equiv x_i - x_{\textrm{true}}(t_i)$. The canonical example in one-dimension (e.g., \citet{press1992-book}) is to assume that the error in our position measurements are Gaussian i.i.d. and are therefore drawn from the following probability distribution
\begin{equation}
\label{gaussian_pdf}
p_g(\epsilon|\sigma_g) = \frac{e^{-\frac{1}{2}\frac{\epsilon^2}{\sigma_g^2}} }{\sigma_g \sqrt{ 2 \pi}},
\end{equation}
where $\sigma_g$ is the standard deviation. This assumption alone places no assumptions on the signal itself, only on the structure of the noise.

The probability of the observed data given model $x(t)$ is
\begin{equation}
\label{max-gaussian}
P = \frac{1}{\sigma \sqrt{2 \pi}} \prod_{i=1}^{N}  \exp \left[ -\frac{1}{2} \left( \frac{x_i - x(t_i)}{\sigma} \right)^2 \right],
\end{equation}
where we have taken $\sigma=\sigma_g$.

Maximizing the probability function in \eqref{max-gaussian} is also the same as minimizing its argument---up to a constant this is the log likelihood, called the penalty function
\begin{equation}
\label{least-squares}
\phi = \frac{1}{N}\sum_{i=1}^{N} \left( \frac{x_i - x(t_i)}{\sigma} \right)^2 .
\end{equation}
Stated in this way it is plain to see that this is the same as asking for the `least-squares' fit of the errors.

\subsection{Smoothing spline penalty function}
\label{smoothing_spline_penalty_function}

The model used here will be the canonical interpolating spline of order $K$ described in section \ref{sec:interpolation}. Of course, we have chosen our knot points such that the model intersects the observations and this certainly maximizes \eqref{max-gaussian} (and minimizes \eqref{least-squares}) because all the errors are zero, but the resulting distribution of errors (a delta function at zero) does not look anything like the assumed Gaussian distribution. Thus, if we want the error distribution that we get out to look like that which we assumed, it is necessary to constrain the problem in some way.

The smoothing spline augments the penalty function of \eqref{least-squares} by adding a global constraint on the $m$-th derivative of the resulting function as in \eqref{smoothing-spline}.
If $\lambda_T \rightarrow 0$ then this reduces to the least-squares fit in \eqref{least-squares}, but if $\lambda_T \rightarrow \infty$ then this forces the model to an $T$-th order polynomial.

To interpret the first term of \eqref{smoothing-spline}, consider a motionless particle at true position $x_0$. Using the $N$ relevant observations $x_i$, the \emph{sample mean} $\bar{x} = \frac{1}{N} \sum x_i$ estimates the particle's position $x_0$. The unbiased \emph{sample variance} estimates the variance of the noise, $\sigma^2$, and is given by $\hat{\sigma}^2=\frac{1}{N-1} \sum (x_i-\bar{x})^2$, the expected value of which is $\left\langle \hat{\sigma}^2 \right\rangle = \left(1-\frac{1}{N}\right)\sigma^2$.

Now consider the opposite extreme where the particle is moving so fast (or the observations are so sparse) that each observation is completely independent of its neighbors. In this case, each observation must be considered separately, so the sample mean at time $t_i$ is just $\bar{x}_i = x_i$ (i.e., we are summing over the single relevant observation). In this scenario we cannot produce a sample variance, because there is only a single relevant observation at time $t_i$.

In practice, the number of relevant observations can be anywhere between $1$ and $N$. Here we use the term \emph{effective sample size}, denoted by $n_{\textrm{eff}}$, to describe the typical number of observations being used to estimate either the particle's position or the variance of the noise at any given time. In this context, the first term of \eqref{smoothing-spline} is proportional to an ensemble of multiple estimates of the sample variance
\begin{equation}
\label{sample_variance}
\hat{\sigma}^2  \equiv \frac{1}{N} \sum_{i=1}^{N} \left( x_i - x(t_i) \right) ^2,
\end{equation}
which is expected to scale as
\begin{equation}
\label{sample_variance_variance}
\left\langle \hat{\sigma}^2 \right\rangle = \left( 1 -  \frac{1}{n^\textrm{var}_{\textrm{eff}}} \right) \sigma^2,
\end{equation}
where $1 < n^\textrm{var}_{\textrm{eff}} \le N$ is our definition of the effective sample size as determined from the sample variance. Revisiting the limiting cases, as $n^\textrm{var}_{\textrm{eff}} \rightarrow N$ the sample variance matches the true variance, but as $n^\textrm{var}_{\textrm{eff}} \rightarrow 1$, the sample variance vanishes. 

There is a very simple physical interpretation for the second term in \eqref{smoothing-spline}. Consider the case where $T=1$ so that the smoothing spline is a constraint on velocity. When averaged over the integration time, the integral produces the root mean square velocity, $u_{\textrm{rms}}$, which means that the second term scales like $u_{\textrm{rms}}^2$. In general, where $x^{(T)}_{\textrm{rms}}$ is the root-mean-square of the $T$-th derivative, this means that $\lambda_T$ scales like
\begin{equation}
\label{lambda}
\lambda_T = \left( 1 -\frac{1}{n^\textrm{var}_{\textrm{eff}}} \right) \frac{1}{ \left(x^{(T)}_{\textrm{rms}}\right)^2}.
\end{equation}
The interpretation of the smoothing spline is therefore that the two terms are balanced by a relative weighting of the sample variance of the noise and mean square of the $T$-th derivative of the physical process. As will be discussed in section \ref{sec:spline_order_tension_order_spectrum}, both $x^{(T)}_{\textrm{rms}}$ and $n^\textrm{var}_{\textrm{eff}}$ can be estimated \emph{a priori} and therefore a good initial estimate for $\lambda_T$ can be made.

\subsection{Smoothing spline maximum likelihood}
\label{sec:maximum_likelihood}

The penalty function for the smoothing spline in \eqref{smoothing-spline} can be restated in terms of maximum likelihood under some conditions (see also chapter 3.8 in \citet{green1994-book}). Assume that in addition to knowing about how the measurement errors are distributed like in \eqref{max-gaussian}, that we also know how the velocity of underlying physical process is distributed. For example, in geophysical turbulence it has been shown that the velocity probability distribution function is like the Laplace distribution \citep{bracco2000-pf}. To recover the smoothing spline, we need to consider the case where the velocity PDF is Gaussian. Stated as maximum likelihood, this means that at \emph{any given instant} (not just the times of observation) we expected the model velocity to look Gaussian. We can discretize the problem by sampling the velocity $Q$ times $t_q = t_1 + q \Delta t_q$, where $\Delta t_q=\frac{t_N-t_1}{Q-1}$ and $q=0..Q-1$. The maximum likelihood is thus stated as
\begin{align}
P =  & \prod^N _{i=1}\frac{1}{\sigma \sqrt{2 \pi}}\exp \left[ -\frac{1}{2} \left( \frac{x_i - x(t_i)}{\sigma} \right)^2 \right] \nonumber \\ &  \cdot  \prod^{Q}_{q=1}\frac{\sqrt{\gamma}}{x^{(T)}_{\textrm{rms}} \sqrt{2 \pi}} \exp \left[  - \frac{\gamma}{2} \left(  \frac{x^{(T)}(t_q)}{x^{(T)}_{\textrm{rms}}} \right)^2 \right],
\label{gaussian-max-likelihood}
\end{align}
which is simply the joint probability of the error distribution from \eqref{max-gaussian} and the velocity distribution of the underlying physical process. We also include parameter $\gamma$ for convenience in order to set the relative weighting between the two distributions, although it could be absorbed into the definition of $x^{(T)}_{\textrm{rms}}$. Writing \eqref{gaussian-max-likelihood} as a penalty function (after converting the product of exponentials into exponentials of sums), we have that
\begin{equation}
-\log P= \frac{1}{2}\sum^N _{i=1}  \left( \frac{x_i - x(t_i)}{\sigma} \right)^2 + \frac{\gamma}{2} \sum^{Q}_{q=1} \left( \frac{x^{(T)}(t_q)}{x^{(T)}_{\textrm{rms}}} \right)^2 + C,
\label{smoothing-spline-log-likelihood}
\end{equation}
where $C$ is a constant. Setting $\gamma=\frac{N}{Q}$ and renormalizing the penalty function by $\frac{2}{N}$ (which has no effect on the location of its minimum), \eqref{smoothing-spline-log-likelihood} can be written as
\begin{equation}
\label{smoothing-spline-pdf}
\phi = \frac{1}{N} \sum^N _{i=1}  \left( \frac{x_i - x(t_i)}{\sigma} \right)^2 + \frac{1}{t_N-t_1} \sum^{Q}_{q=1}  \left(  \frac{x^{(T)}(t_q)}{x^{(T)}_{\textrm{rms}}} \right)^2 \Delta t_q.
\end{equation}
Apart from the discretization of the integral, \eqref{smoothing-spline-pdf} is the same as the penalty function for a smoothing spline \eqref{smoothing-spline}.

There is an important special case when tension is applied at the same order as the spline, $T=S$. In this case the spline is piecewise constant for $x^{(T)}$ with exactly $N-T$ unique values. The parameter $\gamma =\frac{N}{N-T}\approx 1$ and \eqref{gaussian-max-likelihood} can be simplified. This case is appealing because only the $N-T$ unique values of the derivative $x^{(T)}$ that can be computed from $N$ data points are being used for tension, which is not the case when $T<S$.

This maximum likelihood perspective shows that adding tension to the penalty function is equivalent to assuming that one of the higher order derivatives in the model (e.g., velocity if $T=1$) is Gaussian. This is therefore making an assumption about the underlying \emph{physical process} of the model. This is in contrast to the first term which is entirely a statement about \emph{measurement noise}.

As an aside, writing the smoothing spline as a maximum-likelihood condition \eqref{gaussian-max-likelihood}, suggests that if the underlying physical process has a non-zero mean value in tension, the fit will not behave as expected. However, smoothing splines can be easily modified to accommodate a mean value in tension, as shown in appendix \ref{sec:numerical_implementation}. 

\subsection{Optimal parameter estimation} \label{sec:optimal_parameter}

For a given choice of $T$ and $\lambda_T$, the minimum solution to \eqref{smoothing-spline} can be found analytically (see \citet{teanby2007-mg} and our appendix~\ref{sec:numerical_implementation}). Once the solution is found the smoothing matrix $\mathbf{S_\lambda}$ is defined as the matrix that takes the observations $\mathbf{x}$ and maps them to their smooth values, $\mathbf{\hat{x}} = \mathbf{S_\lambda} \mathbf{x}$.

The free parameter $\lambda_T$ is a relative weighting between the two terms in \eqref{smoothing-spline} and choosing its optimal value can be done by minimizing the expected mean square error \citep{craven1979-nm},
\begin{align}
\label{MSE}
    \textrm{MSE}(\lambda) =& \frac{1}{N} || \left( \mathbf{S_\lambda} - I \right) \mathbf{x} ||^2 + \frac{2 \sigma^2}{N}  \Tr \mathbf{S_\lambda} - \sigma^2,
\end{align}
where $||\cdot||^2$ is the Euclidean norm and $\Tr$ indicates the trace.

It is worth noting that a fair amount of the literature on smoothing splines is devoted to minimizing the mean square error when the variance, $\sigma^2$, is \emph{not} known. For example, \citet{craven1979-nm} and \citet{wahba1978-jrss-b} use cross-validation to estimate $\sigma$ and minimize the mean square error. Recent work comparing different estimators shows that no single technique appears to be optimal \citep{lee2003-csda}. For our application however, the errors in GPS data can be relatively easily established, as shown in section \ref{sec:drifter_data_set}.

\begin{table*}[ht]
\caption{68th percentile range of increase in mean square error from the optimal fit}
\label{optimal_T}
\centering
\begin{tabular}{l *{5}{l}}
\toprule & \multicolumn{5}{c}{T} \\ 
\cmidrule(lr){2-6} 
S & 1 & 2 & 3 & 4 & 5 \\ \midrule 
1 & 33.8-80.3\% & & & & \\ 
2 & 14.0-75.1\% & 0.8-12.1\% & & & \\ 
3 & 17.1-77.5\% & 1.0-13.1\% & 0.0-4.5\% & & \\ 
4 & 22.8-81.9\% & 1.0-14.5\% & 0.0-4.6\% & 0.0-6.3\% & \\ 
5 & 27.6-91.4\% & 0.8-15.4\% & 0.0-4.6\% & 0.0-6.1\% & 0.0-12.8\% \\ 
 \bottomrule 
\end{tabular} 
\end{table*}

The mean square error in \eqref{MSE} is a combination of the sample variance and the variance of the mean. As already discussed in the context of the penalty function $\phi$ in section \ref{smoothing_spline_penalty_function}, the first term in \eqref{MSE} is an ensemble of sample variances, and therefore by combining \eqref{sample_variance},  \eqref{sample_variance_variance} and \eqref{MSE} we obtain

\begin{equation}
\label{dof_var}
    \left(1-\frac{1}{n_{\textrm{eff}}^{\textrm{var}}} \right)\sigma^2 = \frac{1}{N} || \left( \mathbf{I} - \mathbf{S_\lambda} \right) \mathbf{x} ||^2.
\end{equation}
The second term in \eqref{MSE} is proportional to twice the squared standard error, i.e., the variance of the sample mean. As discussed in \citet{teanby2007-mg}, the quantity $\mathbf{S_\lambda} \Sigma$ is the covariance matrix with the squared standard error along the diagonal and thus the mean squared standard error is given by $\frac{1}{N} \Tr \left( \mathbf{S_\lambda} \Sigma \right)$. The variance of the sample mean is known to scale inversely with the number of samples being used to estimate the mean. Thus, we use this to define the effective sample size of the variance of the mean, $n_{\textrm{eff}}^{\textrm{SE}}$ with
\begin{equation}
\label{dof_se}
    \frac{\sigma^2}{n_{\textrm{eff}}^{\textrm{SE}}} = \frac{1}{N} \Tr \left( \mathbf{S_\lambda} \Sigma \right).
\end{equation}
Taking the measures of effective sample size as functions of $\lambda$, the mean square error can be expressed by combining \eqref{MSE}--\eqref{dof_se} such that
\begin{equation}
    \textrm{MSE}(\lambda) = 2\frac{\sigma^2}{n_{\textrm{eff}}^{\textrm{SE}}} - \frac{\sigma^2}{n_{\textrm{eff}}^{\textrm{var}}}.
\end{equation}
If one assumes that $ n_{\textrm{eff}}^{\textrm{var}} = n_{\textrm{eff}}^{\textrm{SE}}$, then the expected mean square error from \eqref{MSE} is equal to $\sigma^2/n_{\textrm{eff}}$. Although not shown here, in an empirical analysis we find that $n_{\textrm{eff}}^{\textrm{var}}$ and $n_{\textrm{eff}}^{\textrm{SE}}$ are approximately equal, although $n_{\textrm{eff}}^{\textrm{var}}$ becomes highly variable when $n_{\textrm{eff}}^{\textrm{SE}}$ approaches 1.

These measures of effective sample size can be used to estimate the value of $\lambda_T$ necessary for optimal tension without minimizing the expected mean square error. Note that the definition of effective sample size used here is related to, but not the same as, the notion of degrees-of-freedom used in \citet{cantoni2002-biom} and references therein.



%
\section{Spline order, tension order, and the spectrum} \label{sec:spline_order_tension_order_spectrum}
%

With a model path \eqref{b-spline-model-intro}, a penalty function \eqref{smoothing-spline}, and a minimization condition \eqref{MSE}, we have all the primary pieces to create a smoothing spline interpolant to the data. However, there are a number of choices that still have to be made. In this section we use synthetically generated data to represent our physical process, and contaminate the process with Gaussian noise as described in section \ref{sec:synthetic_data}. We use this synthetically generated data to test our ability to recover the signal and examine the effects of changing the spline and tension order on the mean square error and the resulting spectrum.

The results of this section are empirical, and it is important to acknowledge upfront that any conclusions reached \emph{may} depend on our particular choice of physical model that generates the signal which has been chosen to resemble the oceanographic data of interest. Nevertheless, our expectation is that the conclusions here are `O(1)' correct, and applicable, at least, to our GPS tracked drifter dataset.

\subsection{Tension degree, $T$} \label{tension_degree}

Given a smoothing spline of degree $S$, the tension in the penalty function \eqref{smoothing-spline} can be applied at any degree $T\leq S$. We use the synthetic data for the three different slopes to empirically establish the relationship between the tension degree, $T$ and the spline degree, $S$.

For $S=1\dots 5$ and all $T\leq S$ we minimize the mean square error against the true values. 
The minimization is performed for 200 ensembles of noise and signal with three slopes ($\omega^{-2}$, $\omega^{-3}$, $\omega^{-4}$) and 5 different strides. For a given slope, stride, and realization of noise, we identify the minimum mean square error across $S$ and $T$ and compare all other values of $S$ and $T$ as a percentage increase relative to that minimum. After aggregating across slopes, strides, and ensembles, the 68\% confidence range is shown in Table \ref{optimal_T}.

The results in Table \ref{optimal_T} show that while setting $T=S$ may not always be optimal, it is never significantly worse than the optimal choice. Thus, for the remainder of the manuscript, we will always take $T=S$. This choice is the same as the special case highlighted in section \ref{sec:smoothing_spline}.

\subsection{Loss of coherence} \label{loss_of_coherence}

\begin{figure}
  \centerline{\includegraphics[width=19pc,angle=0]{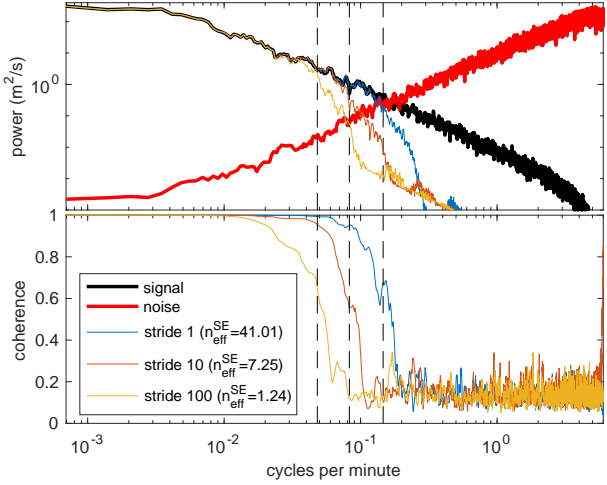}}
  \caption{The upper panel shows the uncontaminated velocity spectrum of the signal (black) and velocity spectrum of the noise (red). The observed signal is the sum of the two. The blue, red, and orange lines show the spectrum of the smoothing spline best fit to the observations with all, 1/10th and 1/100th the data, respectively. The bottom panel shows the coherence between the smoothed signals and the true signal.}
  \label{synthetic_process_and_spectrum}
\end{figure}

The loss-of-coherence defines the time scale below which the smoothing spline is not providing useful information. A reasonable hypothesis is that this scale is related to the effective sample size, $n_{\textrm{eff}}$ because the effective sample size indicates how many points are being used to estimate the true value. Therefore the loss-of-coherence occurs at the \emph{effective Nyquist} which we define as
\begin{equation}
\label{effective-nyquist}
    f_s^{\textrm{eff}} \equiv \frac{1}{2 n_{\textrm{eff}} \Delta t}.
\end{equation}
In practice, we use $n^\textrm{SE}_\textrm{eff}$ because it is less variable than $n^\textrm{var}_\textrm{eff}$ for values near 1 and is the more direct measure of how many points are being used to estimate the model path. 
Fig. \ref{synthetic_process_and_spectrum} shows the power spectrum and coherence of optimal tension fits for three different strides of the data. In all three cases \eqref{effective-nyquist} indicates almost exactly where the coherence drops below 0.5.

\subsection{Reduced spline coefficients} \label{reduced_coefficients}

One practical consideration when working with large datasets is that the computational cost of creating the spline fit may be limited by the rate of solving for the spline coefficients. It is therefore beneficial to reduce knot points (and therefore total splines) where possible. A reasonable hypothesis is to suppose that when the effective sample size is large, as measured by \eqref{dof_se}, that we may be able to avoid placing a knot point at every data point---essentially `skipping' data points.

To test this idea, we find the optimal fit over a range of different strides (which varies the effective sample size) and increase the number of knot points that are skipped until the mean square error starts to rise. We find that we can safely skip $\textrm{max}(1,\textrm{floor}(2n_{\textrm{eff}}/3))$ knot points without sacrificing any precision. In fact, as can be seen in Table \ref{fit_results_gaussian}, in some cases the optimal mean square error improves with fewer knot points. The `full dof' column indicates a fit where one knot point is created for every observation point, whereas the `reduced dof' indicates a fit where the number of knot points is reduced.

\begin{table}[ht]
\caption{Mean square error and effective sample size for a range of strides and smoothing spline methods. 
}
\label{fit_results_gaussian}
\centering
\begin{tabular}{r r p{1.15cm} | p{1cm}p{1cm}p{1cm}p{1cm}} stride & $n_\textrm{eff}$ & optimal mse & reduced dof & blind initial & expected mse \\ \hline \hline 
$\omega^{-2}$ &&&&&  \\ \hline 
1 & 8.6 & 11.5 m$^2$ &  0.1\%  &  56.4\%  &  7.4\%  \\ 
2 & 4.9 & 20.4 m$^2$ &  0.0\%  &  36.3\%  &  2.8\%  \\ 
4 & 2.9 & 34.2 m$^2$ &  0.1\%  &  20.0\%  &  1.7\%  \\ 
8 & 1.7 & 55.9 m$^2$ &  0.0\%  &  5.6\%  &  1.0\%  \\ 
16 & 1.2 & 81.8 m$^2$ &  0.0\%  &  3.6\%  &  0.5\%  \\ 
$\omega^{-3}$ &&&&&  \\ \hline 
1 & 12.5 & 7.64 m$^2$ &  -0.1\%  &  38.6\%  &  6.4\%  \\ 
2 & 7.1 & 13.4 m$^2$ &  -0.1\%  &  20.4\%  &  3.5\%  \\ 
4 & 4.1 & 23.5 m$^2$ &  -0.0\%  &  9.8\%  &  2.2\%  \\ 
8 & 2.3 & 41.8 m$^2$ &  0.0\%  &  1.7\%  &  1.2\%  \\ 
16 & 1.4 & 67.9 m$^2$ &  0.0\%  &  9.6\%  &  0.6\%  \\ 
$\omega^{-4}$ &&&&&  \\ \hline 
1 & 15.6 & 5.69 m$^2$ &  -0.1\%  &  33.8\%  &  7.9\%  \\ 
2 & 9.0 & 10.5 m$^2$ &  -0.1\%  &  18.6\%  &  5.1\%  \\ 
4 & 5.0 & 18.6 m$^2$ &  -0.0\%  &  8.6\%  &  2.4\%  \\ 
8 & 2.8 & 33.2 m$^2$ &  0.0\%  &  3.2\%  &  1.5\%  \\ 
16 & 1.6 & 57.6 m$^2$ &  0.0\%  &  15.4\%  &  0.8\%  \\ 
\end{tabular} 
\end{table}

This means that when handling large datasets, we can reduce the number of splines being used if the effective sample size is large, and we can simply `chunk' the data (split into multiple independent pieces) when the effective sample size is small.

\subsection{Interpolation condition} \label{interpolation_condition}

To estimate the value of $\lambda_T$ from \eqref{lambda}, we require an estimate of the mean square value of a derivative of the process, $x_{\textrm{rms}}^{(T)}$ as well as an estimate of the effective sample size, $n_{\textrm{eff}}$. Assuming one can make an estimate of $x_{\textrm{rms}}^{(m)}$ from the signal (see appendix \ref{sec:variance_estimate}), we just need a method for estimating the effective sample size.

We argue that the effective sample size should vary based on the relative size of the measurement errors to the speed of motion. For example, if the position errors are only $1$ meter, but a particle typically travels $10$ meters between measurements, then it is hardly justifiable to increase the tension so that the smoothing spline misses the observation points by $1$ meter. There is not enough statistical evidence to suggest that the particle didn't go right through the observation point. On the other hand, if the position errors are $1$ meter, but the particle typically travels $10$ centimeters between measurements, nearby measurements provide more information about the particle's true position during that time, so our estimate of the particle's true position is closer to a mean of the nearby observations.

This idea can be made more rigorous by noting that one would consider change in position, $\Delta x$, statistically significant if it exceeds the position errors $\sigma$ by some factor.  Assuming the physical process has a characteristic velocity scale, $u_{\textrm{rms}}$, we use this concept to define $\Gamma$ as
\begin{equation}
\label{gamma_def}
\Gamma \equiv \frac{\sigma}{u_{\textrm{rms}}\Delta t},
\end{equation}
where $\Delta t$ is the typical time between observations. This argument suggests that the effective sample size should be proportional to $\Gamma$, i.e.,
\begin{equation}
\label{gamma_equation}
n_{\textrm{eff}}^\Gamma = \max\left(1,C \cdot \Gamma^m\right)
\end{equation}
where $C$ and $m$ are unknown constants, and we prevent the effective sample size from dropping below 1. Intuitively this means that as long as the particle does not move too far between observations, nearby observations help to estimate the true position of the particle.

\begin{figure}
  \centerline{\includegraphics[width=19pc,angle=0]{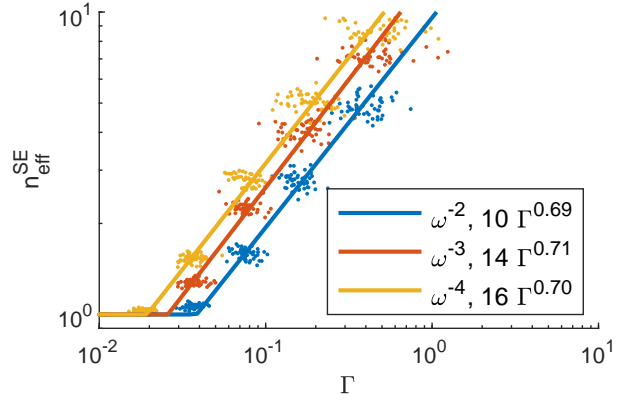}}
  \caption{Effective sample size from the standard error vs $\Gamma$}
  \label{dofVsGamma}
\end{figure}

To test the relationship between $\Gamma$ and the effective sample size, we compute the optimal smoothing spline for a range of values of $\Gamma$ (created by sub-sampling the signal) for the three different slopes ($\omega^{-2}$, $\omega^{-3}$, $\omega^{-4}$). The value $n_{\textrm{eff}}^\textrm{SE}$ is computed from the optimal solution for 50 ensembles and shown in Fig. \ref{dofVsGamma}. The fits are remarkably good, but depend on the slope of process. Processes with shallower slopes (rougher trajectories) provide a smaller effective sample size for a given value of $\Gamma$. 

Using the interpolation condition $\Gamma$ to estimate the effective sample size, we set $n_{\textrm{eff}}^\Gamma = 14 \cdot \Gamma^{0.71}$, the empirically determined best fit for slope $\omega^{-3}$.  For all spline fits then, we use
\begin{equation}
\label{lambda_initial_guess}
\lambda^{\textrm{initial}}_T = \left( 1 -\frac{1}{n_{\textrm{eff}}^\Gamma}\right) \frac{1}{ \left(x^{(T)}_{\textrm{rms}}\right)^2  }
\end{equation}
as an initial estimate for the optimal smoothing parameter where both $x^{(T)}_{\textrm{rms}}$ in \eqref{lambda_initial_guess} and $u_\textrm{rms}$ in \eqref{gamma_def} are estimated using the method described in appendix \ref{sec:variance_estimate}.

The scaling law for $n_{\textrm{eff}}^\Gamma$ can be found analytically.
Let the position observations be given by $x_i$ where
\begin{equation}
x_i = u_\textrm{rms} i \Delta t + \epsilon_i \textrm{ where } \epsilon_i = \mathcal{N}(0,\sigma).
\end{equation}
If the effective sample size is $\left\langle n \right\rangle$, then the particle changes position by $\left\langle n \right\rangle u_\textrm{rms} \Delta t$ between samples. Applying the two-sample $z$-test, two positions will be considered different for $z>z_\textrm{min}$ where
\begin{equation}
z= \frac{\left\langle n \right\rangle u_\textrm{rms} \Delta t}{\sqrt{\frac{\sigma^2}{\left\langle n \right\rangle} + \frac{\sigma^2}{\left\langle n \right\rangle} }} \quad
\Rightarrow \quad
\label{gamma_analytical}
\left\langle n \right\rangle = \left( \frac{z \sigma \sqrt{2}}{u_\textrm{rms} \Delta t} \right)^{\frac{2}{3}}.
\end{equation}
The power law in \eqref{gamma_analytical} matches the empirically derived power laws shown in Fig. \ref{dofVsGamma} and suggests that $m$ in \eqref{gamma_equation} should be $m=2/3$. This also suggests that the coefficient $C$ in \eqref{gamma_equation} can be related to $z$, a measure of statistical significance.

\subsection{Optimal fits} \label{optimal_fits}

Table \ref{fit_results_gaussian} summarizes the key results of this section by applying a smoothing spline with with $S=3$ to the 200 ensembles of the noise and signal with three different slope ($\omega^{-2}$, $\omega^{-3}$, $\omega^{-4}$) and five different strides. The second and third columns show the effective sample size and average mean square error when the smoothing spline is applied using the true values to minimize the mean square error---this is the lower bound. The fourth column shows average increase in mean square error when reducing the number of spline coefficient as documented in section \ref{reduced_coefficients}. There is almost no change in mean square error and therefore all subsequent methods (whether blind or unblind) use this technique. The fifth column uses \eqref{lambda_initial_guess} from section \ref{interpolation_condition} to provide a (blind) initial guess of the tension parameter. Here the results are mixed---a typical increase in mean square error is about 30-50\% when the effective sample size is large. While this might seem large, this is a small fraction of the total variance of the noise, e.g., an optimal mean square error of $6$ m$^2$ increase to $8$ m$^2$ when the total variance is $100$ m$^2$. When the data sets are small (and computation time is not a limiting factor), nearly optimal fits can be found using \eqref{MSE}, as shown in the last column of the table.

\subsection{Numerical implementation}

The numerical implementation of the methods in this section are available in the \texttt{SmoothingSpline} class which subclasses \texttt{BSpline}. This class is initialized with three required parameters: a set of data points $(t_i,x_i)$ and a distribution (specifically a normal distribution for the results in this section). The initial value of $\lambda_T$ is chosen using \eqref{lambda_initial_guess}. The \texttt{SmoothingSpline} class implements a \texttt{.minimize()} method which takes any function of the spline as an argument (such as \eqref{MSE}), and minimizes the function by varying $\lambda_T$.

\section{Bivariate smoothing splines and stationarity}
\label{sec:bivariate}

Up to this point we have considered univariate data, $(t_i, x_i)$, but GPS position data is fundamentally bivariate. The term `bivariate' in the context of splines is often used to denote splines defined on two independent variables---however, in this context we define bivariate to mean two dependent variables (e.g., $x$ and $y$) and one independent variable (e.g., $t$).

The trivial approach to working with such bivariate data is to treat each direction independently---i.e., minimize $\lambda^x_T$ and $\lambda^y_T$ independently of each other. However, it is often the case that the underlying physical process is isotropic. In the context of the maximum likelihood formulation of smoothing splines \eqref{smoothing-spline-pdf}, this means that we expect $x^{(T)}_{\textrm{rms}}$ (the rms value of the tensioned variable) to be the same in all directions (invariant under rotation). This however does \emph{not} mean that $\lambda_x$ should necessarily equal $\lambda_y$. To be explicit, if 
\begin{equation}
\label{lambda_x}
\begin{split}
\lambda^x_T = \left( 1 - \frac{1}{n^x_{\textrm{eff}}} \right) \frac{1}{ \left(x^{(T)}_{\textrm{rms}}\right)^2}, \\
\lambda^y_T = \left( 1 - \frac{1}{n^y_{\textrm{eff}}} \right) \frac{1}{ \left(y^{(T)}_{\textrm{rms}}\right)^2},
\end{split}
\end{equation}
then even if $x^{(T)}_{\textrm{rms}} = y^{(T)}_{\textrm{rms}}$, the effective sample sizes $n^x_{\textrm{eff}}$ and $n^y_{\textrm{eff}}$ will not necessarily be equal if there is any mean velocity because, as shown in section \ref{interpolation_condition}, the effective sample size depends on velocity.

Therefore to assume isotropy in $\lambda_T$ and use a bivariate smoothing spline, the mean velocity from the underlying process must be removed. What qualifies as mean and fluctuation rarely has a clear answer, but a reasonable option is letting a polynomial of degree $T+1$ define the mean. This has the added benefit of removing a constant non-zero tension value, which as shown in section \ref{sec:maximum_likelihood}, changes the problem formulation. 

It is worth noting that it is not actually isotropy that requires removing the mean velocity, but in fact stationarity. The effective sample size is shown to be dependent on rms velocity, so if the velocity varies in time, then the optimal effective sample size will need to vary as well. This means that not only do smoothing splines require stationarity in the tensioned variable $x^{(T)}$ as shown in section \ref{sec:maximum_likelihood}, but they also require stationarity in the velocity $x^{(1)}$ to be effective. This last requirement can be solved by either removing the mean (as we have suggested), or segmenting observations into pseudo-stationary chunks.

\subsection{Assessing errors}
\label{sec:errors_wit_mean}

Removing the mean or some other low-passed version of the data means that the total smoothing matrix will be some combination of the low-passed and high-passed smoothing matrices. Once this matrix is computed, it can be used to compute the standard errors.

We first create a low pass filter to capture the \emph{mean} component of the flow using a simple polynomial fit,
\begin{equation}
\bar{\mathbf{x}} = \bar{\mathbf{S}} \mathbf{x}
\end{equation}
and then define the residual as our stationary part,
\begin{equation}
\mathbf{x}^\prime \equiv \mathbf{x} - \bar{\mathbf{x}}.
\end{equation}
We now compute the smoothing spline as usual on the residual,
\begin{equation}
\mathbf{x}^\prime_\lambda = \mathbf{S}_\lambda \mathbf{x}^\prime
\end{equation}
So the total, smoothed path is
\begin{align}
\hat{\mathbf{x}} =& \bar{\mathbf{x}} + \mathbf{x}^\prime_\lambda = \bar{\mathbf{S}} \mathbf{x} + \mathbf{S}_\lambda \left( \mathbf{x} - \bar{\mathbf{S}} \mathbf{x} \right) = \left(\bar{\mathbf{S}} + \mathbf{S}_\lambda - \mathbf{S}_\lambda \bar{\mathbf{S}}\right)\mathbf{x} \nonumber \\
\equiv& \mathbf{S}_T \mathbf{x}
\end{align}
From this we can compute the covariance matrix and the standard error.

\subsection{Numerical implementation}

The \texttt{BivariateSmoothingSpline} class is initialized with data $(t_i, x_i, y_i)$ and a distribution. For a spline of degree $S=T$, a spline of degree $S+1$ is used to remove the mean in each direction. In the case of a normal distribution, this is simply a least squares polynomial fit. By assumption, the residual data ($\mathbf{x}^\prime$, $\mathbf{y}^\prime$ in the notation above) is stationary and isotropic, so the tension parameter $\lambda_T$ is applied equally to spline fits in the two directions. Minimization is performed on the sum of the expected mean square error in both directions.

\section{GPS data set}
\label{sec:drifter_data_set}


The primary dataset considered here will be nine surface drifters that were deployed in the Sargasso Sea in the summer of 2011 \citep{shcherbina2015-bams}. 
In the past, such drifters used the Argos positioning system which has significantly poorer temporal coverage and position accuracy \citep{elipot2016-jgr}, but recently the majority of surface drifters have employed GPS receivers and transmitted their data back through Argos or Iridium satellites.

The GPS receiver sits on the surface drifter and collects position data, but because of atmospheric conditions or ocean waves, the receivers are sometimes unable to obtain a position, or when they do, it is highly inaccurate. Thus, despite nominal accuracies of a few meters, it is often the case that some positions are off by more than 1000 meters, as can be seen in Fig. \ref{gpsfit}. Applying a smoothing spline fit using the methodology in section \ref{sec:smoothing_spline} produces an extremely poor fit, with clear overshoots to bad data points.


%
\subsection{GPS error distribution}
\label{gps_position_errors}
%

We characterize the GPS errors by considering data from a motionless GPS receiver allowed to run for 12 hours. The specific GPS receiver used for this test was not the same as the one used for the drifters (because it was no longer available) but should produce errors similar enough for this analysis.

\begin{figure}
  \centerline{\includegraphics[width=15pc,angle=0]{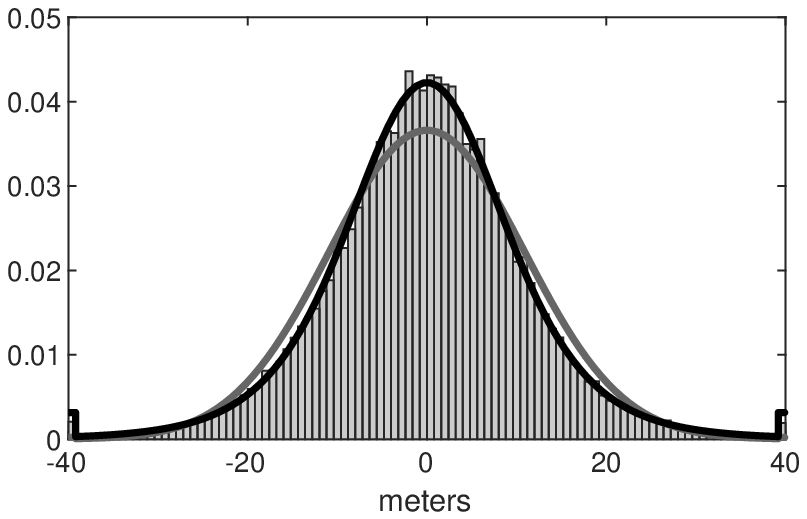}}
  \centerline{\includegraphics[width=15pc,angle=0]{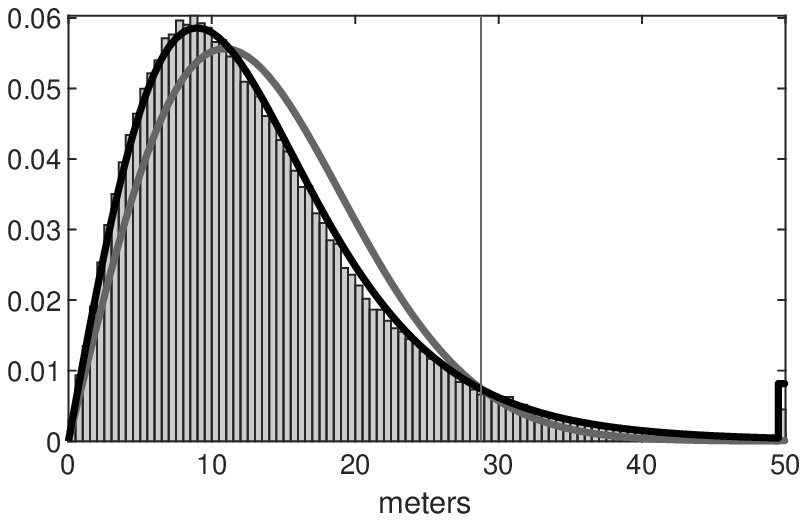}} 
  \caption{The top panel shows the position error distribution of the motionless GPS. The gray/black lines are the best fit Gaussian/$t$-distributions respectively. The bottom panel shows the distance error distribution with the corresponding expected distributions from the Gaussian and $t$-distribution. The vertical line in the bottom panel shows the 95\% error of the $t$-distribution.}
  \label{motionless_error}
\end{figure}

The position recorded by the motionless GPS are assumed to have isotropic errors with mean zero, which means that the positions themselves are the errors. The probability distribution function (PDF) of the combined $x$ and $y$ position errors are shown in Fig. \ref{motionless_error}.

The error distribution is first fit to a zero-mean Gaussian PDF \eqref{gaussian_pdf}.
The maximum likelihood fit is found by simply computing the standard deviation of the sample, which is found to be $\sigma \approx 10$ meters and shown as the gray line in Fig. \ref{motionless_error}. However, it is clear the error distribution shows much longer tails than the Gaussian PDF.

The Student $t$-distribution is a generalization of the Gaussian that produces longer tails and is defined as 
\begin{equation}
\label{student_pdf}
p_s\left(\epsilon |\nu,\sigma_s^2\right) = \frac{\Gamma\left( \frac{\nu + 1}{2} \right)}{\sigma_s \sqrt{\nu \pi} \Gamma\left(\frac{\nu}{2}\right)} \left( 1 + \frac{\epsilon^2}{\sigma_s^2 \nu} \right)^{-\frac{\nu+1}{2}},
\end{equation}
where the $\sigma_s$ parameter scales the distribution width and the $\nu$ parameter sets the number of degrees of freedom. The variance is $\sigma^2=\sigma_s^2 \frac{\nu}{\nu-2}$ and only exists for $\nu > 2$. The $t$-distribution is equivalent to the Gaussian distribution when $\nu \rightarrow \infty$. We find the best fit $t$-distribution to the data by minimizing the Anderson-Darling test. The best fit with parameters $\sigma_s \approx 8.5$ meters and $\nu \approx 4.5$ is shown as the black line in Fig. \ref{motionless_error}. Different choices in GPS receivers and using the Kolmogorov-Smirnoff test results in very similar parameters, i.e., $\sigma_s\approx8-10$ meters and $\nu\approx4-6$.

The \emph{position} error distributions also imply a combined \emph{distance} error distribution by computing $\epsilon_d = \sqrt{\epsilon_x^2 + \epsilon_y^2}$ and is shown in the lower panel of Fig. \ref{motionless_error}. For two independent Gaussian distributions this results in a Rayleigh distribution,
\begin{equation}
\label{rayleigh_pdf}
p_r(\epsilon_d|\sigma_g) = \frac{\epsilon_d}{\sigma_g^2 } e^{-\frac{1}{2}\frac{\epsilon_d^2}{\sigma_g^2}}.
\end{equation}
The distance distribution for two $t$-distributions is computed numerically and is shown in the bottom panel of Fig. \ref{motionless_error} on top of the actual distance errors. Approximately 95\% of distance errors are within $30$ meters.

\begin{figure}
  \centerline{\includegraphics[width=15pc,angle=0]{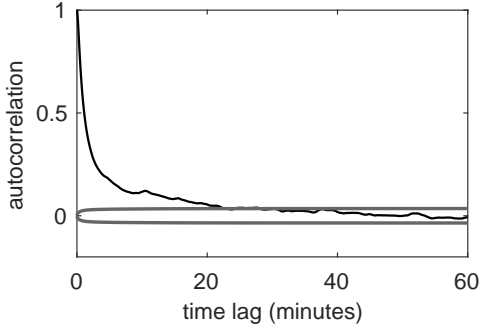}}
  \caption{The autocorrelation function of the GPS positioning error with 99\% confidence intervals shown in gray. The correlation at drifter sampling period of 30 minutes is indistinguishable from zero.} 
  \label{gps_autocorrelation}
\end{figure}

Fig. \ref{gps_autocorrelation} shows the autocorrelation function of the GPS position errors. We find a rough empirical fit to be $\rho(\tau)=\exp\left(\max(-\tau/t_0,-\tau/t_1-1.35)\right)$ where $t_0=100$ seconds and $t_1=760$ seconds, which reflects an initially rapid fall off in correlation, followed by a slower decline. The smallest sampling interval of the GPS drifters in question is 30 minutes and therefore it is safe to assume the errors are uncorrelated for our purposes. Although the drifter sampling rate allows us to avoid further discussion of the autocorrelation function of GPS errors, accounting for autocorrelation is a relatively easy extension (and in fact, already implemented in the code).

The smoothing spline algorithms described in section \ref{sec:smoothing_spline} are modified to use the $t$-distribution as described in section \ref{sec:irls}. Table \ref{fit_results_tdistribution} shows that the conclusions reached for Gaussian data in section \ref{sec:smoothing_spline} still apply with $t$-distributed data.

\section{Minimization with Outliers}
\label{sec:outliers}

\begin{figure*}[t]
  \centering
    \begin{minipage}{0.30\textwidth}
        \centering
        \includegraphics[width=1.0\textwidth]{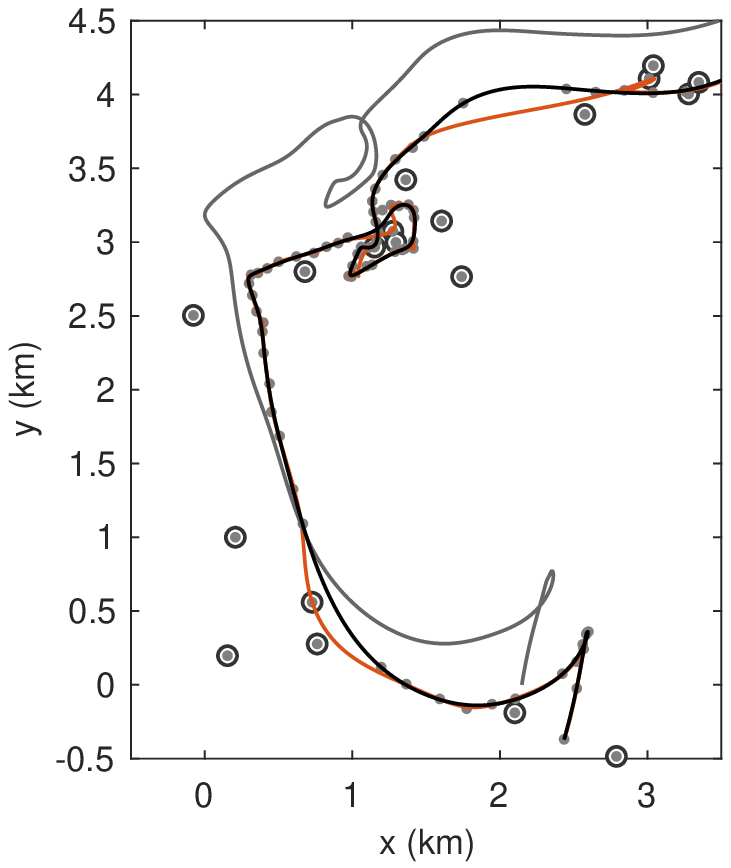} 
    \end{minipage}\hfill
    \begin{minipage}{0.7\textwidth}
        \centering
        \includegraphics[width=1.0\textwidth]{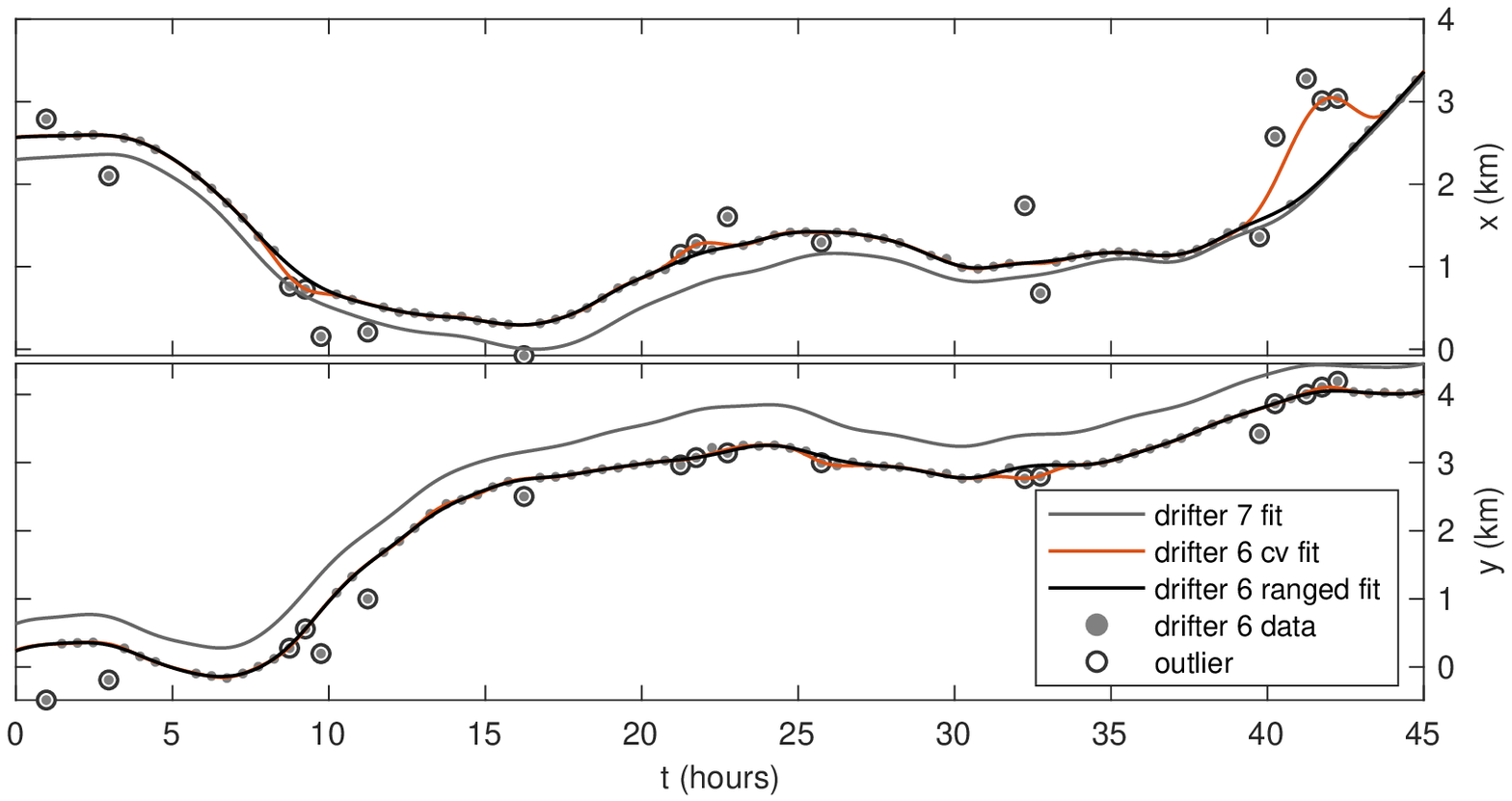} 
    \end{minipage}
   \caption{GPS position data for a 40 hour window from drifter 6. The points are the recorded positions and the black line is the optimal fit using the ranged expected mean square error. Data points with less than 0.01\% chance of occurring are highlighted and deemed outliers. The light grey line is the is optimal smoothing spline fit for drifter 7, which has no apparent outliers and was released a few hundred meters from drifter 6. The orange line is the smoothing spline fit assuming t-distributed errors, but using cross-validation to minimize $\lambda_T$}
  \label{gpsfit}
\end{figure*}

The goal here is to find a smooth solution in the presence of outliers---points that do not appear to be of the known error distribution for the GPS receiver shown in section \ref{gps_position_errors}. These points are obviously problematic as can be seen in Fig. \ref{gpsfit}, where individual data points jump hundreds of meters and even several kilometers away from its neighbors. Errors of this size are inconsistent with the noise analysis of the preceding section, so the goal here is to find a model path $x(t)$ robust to this uncharacterized noise. What makes outliers `obvious' to the eye is that they appear as unexpectedly large motions, inconsistent with most of the other motion for that path. In this sense, the smoothing spline formulation is a good one as it assumes the motion at some order (e.g., acceleration) is Gaussian, as shown in section \ref{sec:maximum_likelihood}. Interestingly, in the nine drifters we are analyzing here, one drifter shows no obvious outliers, suggesting the issue may be related to how the antenna is configured. This particular drifter serves as a useful point of comparison.

\begin{table}[ht]
\caption{Same as Table \ref{fit_results_gaussian}, but with noise following a $t$ distribution.}
\label{fit_results_tdistribution}
\centering
\begin{tabular}{r r p{1.15cm} | p{1cm}p{1cm}p{1cm}p{1cm}} stride & $n_\textrm{eff}$ & optimal mse & reduced dof & blind initial & expected mse \\ \hline \hline 
$\omega^{-2}$ &&&&&  \\ \hline 
1 & 8.2 & 11.8 m$^2$ &  0.3\%  &  66.7\%  &  7.7\%  \\ 
2 & 4.7 & 20.9 m$^2$ &  0.3\%  &  47.3\%  &  6.6\%  \\ 
4 & 2.8 & 38.0 m$^2$ &  0.1\%  &  24.2\%  &  4.4\%  \\ 
8 & 1.6 & 66.3 m$^2$ &  0.0\%  &  8.2\%  &  9.3\%  \\ 
16 & 1.2 & 101. m$^2$ &  0.0\%  &  8.1\%  &  3.7\%  \\ 
$\omega^{-3}$ &&&&&  \\ \hline 
1 & 12.1 & 7.51 m$^2$ &  -0.1\%  &  36.2\%  &  8.8\%  \\ 
2 & 6.8 & 13.4 m$^2$ &  -0.1\%  &  22.8\%  &  7.0\%  \\ 
4 & 3.9 & 26.0 m$^2$ &  -0.0\%  &  11.5\%  &  3.8\%  \\ 
8 & 2.2 & 47.5 m$^2$ &  0.0\%  &  2.2\%  &  3.2\%  \\ 
16 & 1.3 & 82.5 m$^2$ &  0.0\%  &  12.6\%  &  8.5\%  \\ 
$\omega^{-4}$ &&&&&  \\ \hline 
1 & 14.9 & 6.01 m$^2$ &  -0.2\%  &  35.3\%  &  9.0\%  \\ 
2 & 8.6 & 10.5 m$^2$ &  -0.2\%  &  24.8\%  &  7.0\%  \\ 
4 & 4.8 & 19.1 m$^2$ &  -0.1\%  &  7.8\%  &  4.6\%  \\ 
8 & 2.7 & 36.4 m$^2$ &  0.0\%  &  3.2\%  &  2.7\%  \\ 
16 & 1.6 & 69.1 m$^2$ &  0.0\%  &  18.9\%  &  11.5\%  \\ 
\end{tabular} 
\end{table}

Minimizing with the expected mean square error \eqref{MSE} produces a fit so poor that it is not worth showing. Because outliers add enormous amounts of variance, the expected mean square error vastly under tensions the spline---essentially chasing every outlier shown in Fig. \ref{gpsfit}. Because some of the noise is uncharacterized, this suggests using a method such as cross-validation might be effective. The orange line in Fig. \ref{gpsfit} uses a smoothing spline fit, assuming Student t-distributed errors, but minimized with cross-validation. This fit performs relatively well, but compared with the drifter 7, it is clear that it still chases some outliers. The goal in this section is to develop a method robust to outliers in cases where we know something about the noise.

The basic problem formulation is as follows: we define a new `robust distribution', $p_{\textrm{robust}}$, that includes the known noise distribution, $p_{\textrm{noise}}$, plus an unknown (or assumed) form of an outlier distribution, $p_{\textrm{outlier}}$,
\begin{equation}
\label{robust_pdf}
    p_{\textrm{robust}}(\epsilon) = (1-\alpha) \cdot p_{\textrm{noise}}(\epsilon) + \alpha \cdot  p_{\textrm{outlier}}(\epsilon).
\end{equation}
We consider
a $t$-distribution for $p_{\textrm{noise}}$ with parameters found from the GPS errors in section \ref{gps_position_errors}. The distribution of $ p_{\textrm{outlier}}$ is also set to be a $t$-distribution, but with $\nu=3$ and $\sigma=50\sigma_\textrm{gps}$ which roughly matches the total variance of the observed outliers. In our tests we varied $\alpha$ from 0 up to $0.25$, approximately the range of observed outliers from the drifter data sets.

Throughout our attempts to smooth the noisy GPS data we tried many different approaches to modifying smoothing splines for robustness to outliers, but ultimately found that enormous gains in accuracy are made by simply discarding outliers while minimizing the expected mean square error \eqref{MSE}. The results of this approach are shown in section \ref{sec:robust_minimization}, but we also document our methodology to reliably estimate the outlier distribution in section \ref{sec:full_tension}.

\subsection{Robust minimization}
\label{sec:robust_minimization}

The whole problem with outliers is that we do not know their distribution, so minimizing the expected mean square error using \eqref{MSE} with the expected variance from the robust distribution defined in \eqref{robust_pdf} cannot possibly work. Outliers add extra variance, and will therefore cause the spline to be under tensioned ($\lambda_T$ too small). The key concept behind our method is to simply exclude the outliers from the calculation of \eqref{MSE}, where outliers are defined as points unlikely to arise with the known noise distribution. The \emph{ranged expected mean square error} thus replaces $\sigma^2$ with,
\begin{equation}
\label{ranged_mse}
    \sigma_\beta^2 = \int_{\textrm{cdf}^{-1}(\beta/2)}^{\textrm{cdf}^{-1}(1-\beta/2)} z^2 p_{\textrm{noise}}(z) \, dz
\end{equation}
and discards all rows (and columns) of $\mathbf{S}_\lambda$ where $\left(\mathbf{S}_\lambda - I\right)\mathbf{x}<\textrm{cdf}^{-1}(\beta/2)$ or $\left(\mathbf{S}_\lambda - I\right)\mathbf{x}>\textrm{cdf}^{-1}(1-\beta/2)$.

To test this approach we generate data as before, but now also let a certain percentage of outliers ($\alpha$) be generated with an outlier distribution following \eqref{robust_pdf}. We consider five different values of $\beta=\left[\frac{1}{50},\frac{1}{100},\frac{1}{200},\frac{1}{400},\frac{1}{800} \right]$ as well as $\beta=0$, which is just \eqref{MSE}. Tests across a number of ensembles with outlier ratios $\alpha=\left[0.0,0.05,0.10,0.25\right]$ we find that $\beta=\frac{1}{100}$ is overall the best choice.

\subsection{Full tension solution and outlier distribution}
\label{sec:full_tension}

The \emph{full tension} solution is defined as the maximum allowable value of $\lambda$ given the known noise distribution. That is, the spline fit is pulled away from the observations so that the distribution of observed errors ($x_i-x(t_i)$) matches the expected distribution $p_{\textrm{noise}}(\epsilon)$. In cases where the effective sample size $n_\textrm{eff}$ is large, the full tension solution will approximately match the optimal (minimal mean square error) solution. In cases where the effective sample size is small, the full tension solution is more akin to a low-pass solution (because increasing $\lambda$ is equivalent to decreasing $x^{(T)}_\textrm{rms}$).

In the simplest case where there are no outliers, the full tension solution can be found by requiring that the sample variance match the variance of $p_{\textrm{noise}}(\epsilon)$. When outliers are present, a more robust method of estimation is required. After some experimentation, we found that the most reliable method of achieving full tension is to minimize the Anderson-Darling test of $p_{\textrm{noise}}(\epsilon)$ on the interquartile range of observed errors. In fact, we found that this method can be used to estimate the outlier distribution and further refine both the full tension solution and the range over which the expected mean square error is computed.

The outlier distribution is estimated in the following fashion. We first assume that the outlier distribution follows a $t$-distribution with $\nu=3$ and that $\alpha<0.5$. If the spline is in full tension, then the observed total variance can be used to find $\sigma_o$ for the outlier distribution. From \eqref{robust_pdf} it follows that,
\begin{equation}
    \textrm{var}_\textrm{total} = (1-\alpha) \textrm{var}_\textrm{noise} + \alpha 3 \sigma_o^2
\end{equation}
which, given some $\alpha$, can be solved for $\sigma_o$. Our method considers 100 different values of $\alpha$ logarithmicaly spaced from $0.01$ to $0.5$ and chooses the value which minimizes the Anderson-Darling test.

With an estimate for $p_{\textrm{robust}}(\epsilon)$, the full tension solution can be refined by now minimizing the Anderson-Darling test of $p_{\textrm{robust}}(\epsilon)$ on the interquartile range of observed errors. This iterative process converges quite quickly on a good estimate for the outlier distribution and the full tension solution.

\subsection{Extension to bivariate data}
\label{sec:robust_bivariate}

The strategies in this section are relatively easily extended to bivariate data. All error distributions are assumed isotropic, and thus the outlier distribution can be estimated by including the errors from both independent directions. The ranged expected mean square error calculation defined in section \ref{sec:robust_minimization} uses the \emph{distance} of the error for its cutoff in order to remain invariant under rotation.

Application of this methodology to one of the GPS drifters (drifter 6) is shown in Fig. \ref{gpsfit}. Although it is impossible to know exactly how well the smoothing spline fit performed, comparison with drifter 7 (with no apparent outliers) suggests that our methodology successfully avoids chasing outliers.

\subsection{Numerical implementation}

The \texttt{GPSSmoothingSpline} inherits from the \texttt{BivariateSmoothingSpline} class and assumes the errors follow the $t$-distribution found in section \ref{gps_position_errors}. The class also projects latitude and longitude using a transverse Mercator projection with the central meridian set to the center of the dataset.

\section{Conclusions}

The methodology manuscript solves our initial problem of finding smoothed, interpolated positions from our noisy GPS drifter dataset with outliers. For signals similar to the Mat\'ern process, we found that
\begin{enumerate}
\item the spline degree $S$ should be set to a value higher than the high frequency slope of the process (section \ref{sec:interpolation})
\item the tension degree $T$ can be set to $T=S$ (section \ref{sec:spline_order_tension_order_spectrum}), and
\item the optimal tension parameter can be estimated \emph{a priori} (also section \ref{sec:spline_order_tension_order_spectrum}).
\end{enumerate}
For the GPS data in particular, there appear to be three key steps for using smoothing splines to achieve these results:
\begin{enumerate}
    \item using a $t$-distribution for the noise (section \ref{sec:drifter_data_set}),
    \item removing the mean velocity to make the bivariate data stationary (section \ref{sec:bivariate}), and
    \item using the ranged expected mean square error for robustness to outliers (section \ref{sec:outliers}).
\end{enumerate}
The effective Nyquist identified in section \ref{loss_of_coherence} indicates that the power spectrum for the GPS drifters resulting from the smoothed fit is valid up to about half the nominal sampling rate.

\section*{Acknowledgments}
Thanks to Miles Sundermeyer whose drifters were used in this analysis. This work was funded by ONR through the Scalable Lateral Mixing and Coherent Turbulence Departmental Research Initiative (LatMix) and National Science Foundation award 1658564.

%
\section*{Appendix A: Numerical implementation}
\label{sec:numerical_implementation}

The B-splines are generated using the algorithm described in \citet{deboor1978-book} with knot points determined by  \eqref{even-knots} and \eqref{odd-knots}. The matrix $\mathbf{X}$ with components $X\indices{^i_m}$ denotes the $m$-th B-spline at time $t_i$. In this notation the column vector $\xi^m$ represents the coefficients of the splines such that positions at time $t_i$ are given by $\hat{x}^i$ where $\hat{x}^i =  X\indices{^i_m} \xi^m$.

The smoothing spline condition given in \eqref{gaussian-max-likelihood} can be augmented to include a nonzero mean tension, $\mu_u$,
\begin{equation}
\phi =  \frac{1}{N} \sum^N _{i=1}\left( \frac{x_i - x(t_i)}{\sigma_i} \right)^2 + \frac{1}{Q} \sum^{Q}_{q=1}  \left(  \frac{u(t_q)-\mu_u}{\sigma_u} \right)^2,
\end{equation}
where we have taken $T=1$ for this calculation. The discretized penalty function is
\begin{equation}
\phi = \left[ \mathbf{x} - \mathbf{X} \mathbf{\xi} \right]^{\textrm{T}} \Sigma^{-1} \left[ \mathbf{x} - \mathbf{X} \mathbf{\xi}\right]
+ \lambda_1 \left[\mathbf{V}\mathbf{\xi} - \mu \right]^{\textrm{T}} \left[ \mathbf{V}\mathbf{\xi} - \mu \right],
\end{equation}
where $\Sigma$ denotes the covariance matrix describing the measurement errors and we absorbed several constants into $\lambda_1$. To find the coefficients that minimize this function, we take the derivative with respect to $\mathbf{\xi}$, set it to zero, and solve for $\mathbf{\xi}$,
\begin{equation}
\label{xi_equation}
\mathbf{\xi} = \left[ \mathbf{X}^{\textrm{T}} \Sigma^{-1} \mathbf{X} + \lambda_1 \mathbf{V}^{\textrm{T}} \mathbf{V} \right]^{-1}   \left[ \mathbf{X}^{\textrm{T}} \Sigma^{-1} \mathbf{x} +  \mu \lambda_1 \mathbf{V}^{\textrm{T}} \mathbf{\iota} \right],
\end{equation}
where $\mathbf{\iota}$ is a vector of $1$s. The operation $\mathbf{V}^{\textrm{T}} \mathbf{\iota}$ essentially integrates the $m$-splines and results in a column vector with the integrated values.

We define the smoothing matrix as the linear operator that takes observations $\mathbf{x}$ to their smoothed values $\mathbf{\hat{x}}$,
\begin{equation}
\mathbf{\hat{x}} = \mathbf{S_\lambda} \mathbf{x}.
\end{equation}
From this definition and \eqref{xi_equation},
\begin{equation}
\label{smoothing-operator}
\mathbf{S_\lambda} \equiv \mathbf{X} \left[ \mathbf{X}^{\textrm{T}} \Sigma^{-1} \mathbf{X} + \lambda_1 \mathbf{V}^{\textrm{T}} \mathbf{V} \right]^{-1} \mathbf{X}^{\textrm{T}} \Sigma^{-1},
\end{equation}
when $\mu=0$.

\section*{B: Iteratively reweighted least squares}
\label{sec:irls}

In practice it is challenging to use the $t$-distribution directly because it does not result in a linear solution for the coefficients as in \eqref{xi_equation}. One method around this issue is to use a search algorithm to directly look for the maximum values. Alternatively, one can use the iteratively reweighted least squares (IRLS) method.

The idea with IRLS is to reweight the coefficients of the Gaussian, $\sigma_g$ in \eqref{gaussian_pdf}, so that the resulting distribution looks like the desired distribution, e.g., \eqref{student_pdf}. Recalling that $\epsilon_i \equiv x_i - x(t_i,\mathbf{\xi})$, the minimization condition that $\frac{d p_g}{d\xi}=0$, implies that
\begin{equation}
\frac{\epsilon_i}{\sigma_g^2} \frac{\partial x(t_i,\mathbf{x})}{\partial \mathbf{\xi}} = 0,
\end{equation}
for the Gaussian distribution, whereas for the $t$-distribution this implies that,
\begin{equation}
 \frac{\epsilon_i}{\sigma_s^2} \frac{\nu+1}{\nu} \left( 1 + \frac{\epsilon_i^2}{\nu \sigma_s^2} \right)^{-1}  \frac{\partial x(t_i,\mathbf{x})}{\partial \mathbf{\xi}}  = 0.
\end{equation}
This means that one can set
\begin{equation}
\sigma_g^2 =   \sigma_s^2 \frac{\nu}{\nu+1} \left( 1 + \frac{\epsilon_i^2}{\nu \sigma_s^2} \right),
\label{sigma_reweighted}
\end{equation}
to get a matching distribution. Of course, this is only true if $\epsilon_i$ is already known, which initially it is not. So the method becomes iterative---one starts with $\epsilon_i$ determined from the Gaussian fit, then determine a new $\epsilon_i$ after reweighting $\sigma_g$. This method iterates until $\sigma_g$ stops changing. We can rewrite \eqref{sigma_reweighted} as a function of $\epsilon_i$,
\begin{equation}
\label{t-weight}
w_s(\epsilon_i) = \sigma_s^2 \frac{\nu  + \frac{\epsilon_i^2}{\sigma_s^2}}{\nu+1}.
\end{equation}

From \eqref{t-weight} it is clear that if $\epsilon_i < \sigma_s$ then it will be reweighted to a smaller value, essentially making the observation point more strongly weighted. On the other hand, if $\epsilon_i > \sigma_s$, then its relative weighting will decrease, and it will be treated more as an outlier.

More generally, the weight function $w(z)$ for a pdf $p(z)$ is found by setting $-\partial_z \log p(z)$ equal to $-\partial_z \log p_g(z)$ of a Gaussian pdf where $w(z)$ replaces $\sigma_g^2$,
and then solving for $w(z)$. The result is that,
\begin{equation}
    \frac{z}{w(z)} = - \frac{\partial_z p}{p}
\quad \Rightarrow \quad
    w(z) = -z \frac{p}{\partial_z p}.
\end{equation}
Note that the same strategy could be used to reshape the pdf of a Gaussian to match the desired distribution, but here we simply match the minimization conditions of the pdfs.

As a point of reference, Tukey's biweight is given by,
\begin{equation}
\psi(z) = 
\begin{cases}
\frac{z}{\sigma_{tb}^2} \left(1-\frac{z^2}{c^2 \sigma_{tb}^2} \right)^2 & |z| < c \cdot \sigma_{tb} \\
0 & \textrm{else},
\end{cases}
\end{equation}
which, as a weight function is,
\begin{equation}
    w_{tb}(\epsilon_i) = \frac{z}{\psi(z)}.
\end{equation}

In a practical sense, the $\Sigma^{-1}$ of \eqref{smoothing-operator} is replaced with the diagonal matrix $W\equiv\textrm{diag}(1/w(\epsilon_i))$ populated with the reweighted values for each observation such that,
\begin{equation}
\label{general-smoothing-operator}
\mathbf{S_\lambda} \equiv \mathbf{X} \left[ \mathbf{X}^{\textrm{T}} W \mathbf{X} + \lambda_1 \mathbf{V}^{\textrm{T}} \mathbf{V} \right]^{-1} \mathbf{X}^{\textrm{T}} W.
\end{equation}
This operator is again used to compute the standard error from the variances,  $\mathbf{S_\lambda} \Sigma$, where the variance is assumed to be $\sigma_s^2 \frac{\nu}{\nu-2}$ for each observation when using a $t$-distribution.

The reality is that the smoothing spline solution \emph{does} depend on the initial value of $w(\epsilon_i)$ used in the IRLS method. That said, we find that for uniform initial weightings (e.g., all values start with the square root of the variance), the differences are not statistically significant from other initial values.

\section*{C: Estimating the variance of the signal}
\label{sec:variance_estimate}

The method in this paper depends on good estimates of the root-mean-square velocity, $u_{\textrm{rms}}$, of the signal in order to determine the effective degrees of freedom, as well as the variance of the tensioned derivative. The approach taken here is to compute the power spectrum of the signal at the derivative of interest, and sum the variance that is statistically significantly greater than the expected variance of the noise.

Given a process observed with values $x_n$ at times $t_n = n \Delta$ where $n=1..N$, we estimate the mean of its $m$-th derivative by performing a least squares fit to the polynomial $\bar{x}_n \equiv p_m t_n^m + p_{m-1} t_n^{m-1} + .. + p_0$. The \emph{detrended} time series is then defined as $\tilde{x}_n \equiv x_n - \bar{x}_n$. The power spectrum of this time series is given by
\begin{equation}
S_{\textrm{signal}}(f_k) = \frac{\Delta}{N} \left\lvert \sum_{n=0}^{N-1} x_n e^{-2\pi i f_k t_n} \right\rvert^2,
\end{equation}
where the frequencies $f_k$ are given by $f_k = \frac{k}{N\Delta}$. 
By Plancherel's theorem,
\begin{equation}
\label{plancherel}
\sum_{k=0}^{N-1} S(f_k) \cdot \frac{1}{N \Delta} = \frac{1}{N \Delta} \sum_{i=0}^{N-1} x_i^2 \Delta.
\end{equation}
The power spectrum of the $m$-th derivative of the process is computed as
\begin{equation}
S_{\textrm{signal}}^{(m)}(f_k) = (2 \pi f_k)^{2m} \cdot S(f_k).
\end{equation}
Note that it is important to detrend the signal prior to computing the derivative because, by assumption, the signal is periodic and has no secular trend.

The noise, $\epsilon_i$, has total variance $\sigma^2 = \frac{1}{N} \sum_{i=1}^{N} \epsilon_i^2$. Because the noise is assumed to be uncorrelated, the variance distributes evenly across all frequency. The spectrum of the noise is therefore
\begin{equation}
S_{\textrm{noise}}(f_k) = \sigma^2 \Delta,
\end{equation}
which immediately can be seen to satisfy Plancherel's theorem \eqref{plancherel}. The $m$-th derivative of the noise has the power spectrum
\begin{equation}
S_{\textrm{noise}}^{(m)}(f_k) = \sigma^2 \Delta (2 \pi f_k)^{2m}.
\end{equation}

The technique used here sums the variance of the signal for a given frequency if it exceeds the expected variance of the noise at the frequency by some threshold. The estimate of power at each frequency follows a $\chi^2$ distribution with 2 degrees-of-freedom, so we choose the threshold based on the 95-th percentile of the expected distribution. And thus,
\begin{equation}
x^{(m)}_{\textrm{std}} = \sum_{k=0}^{N-1} S^{(m)}_{\textrm{signal}}(f_k) \cdot \left( S^{(m)}_{\textrm{signal}}(f_k) > q S_{\textrm{noise}}^{(m)}(f_k) \right) \cdot \frac{1}{N \Delta},
\end{equation}
where $q\approx 20$ for the 95-percent confidence.

\bibliographystyle{ametsoc2014}

\end{document}